\documentclass[12pt]{article}
\usepackage{epsf}
\def\be{\begin{equation}}
\def\ee{\end{equation}}
\def\bea{\begin{eqnarray}}
\def\eea{\end{eqnarray}}

\begin{document}
\begin{titlepage}
\begin{center}
{\Large \bf William I. Fine Theoretical Physics Institute \\
University of Minnesota \\}
\end{center}
\vspace{0.2in}
\begin{flushright}
FTPI-MINN-17/11 \\
UMN-TH-3627/17 \\
April 2017 \\
\end{flushright}
\vspace{0.3in}
\begin{center}
{\Large \bf Loops with heavy particles in multi Higgs production amplitudes.
\\}
\vspace{0.2in}
{\bf  M.B. Voloshin  \\ }
William I. Fine Theoretical Physics Institute, University of
Minnesota,\\ Minneapolis, MN 55455, USA \\
School of Physics and Astronomy, University of Minnesota, Minneapolis, MN 55455, USA \\ and \\
Institute of Theoretical and Experimental Physics, Moscow, 117218, Russia
\\[0.2in]

\end{center}

\vspace{0.2in}

\begin{abstract}
In view of a recently renewed interest to production of multiple Higgs bosons the amplitude for such process at the threshold of $n$ particles is considered. An explicit calculation is done for the loop corrections to the amplitude arising from interaction with a heavy fermion (e.g. top quark) and also with a heavy scalar. It is shown that such corrections generally break the scaling dependence on the number $n$ and the Higgs self coupling, which dependence is known from the past studies of models with one field. The correction due to the top quark loop is also found to be numerically large and exceeding that from the self coupling up to very high $n$.

\end{abstract}
\end{titlepage}

Multiple production of weakly interacting particles is naturally suppressed by a corresponding high power of small coupling constant. However the number of graphs describing the production amplitude also grows factorially so that the yield of, say $n$ Higgs bosons, at sufficiently high energy contains the factor $n! \lambda^n$ that hints at the total cross section possibly becoming large at large $n$, $n > 1/\lambda$ as the factorial $n!$ overcomes the high power of the small Higgs coupling $\lambda$. The tantalizing prospect of finding a large yield in multiparticle weak interaction processes had stimulated great interest and intensive studies in the early 1990's (a review can be found in Ref.~\cite{mv94}). The general conclusion from  that past activity, although not entirely certain, was that the seemingly large probability at large $n$, is likely a ``mirage" caused by extrapolation of low $n$ results, and that the actual probability at large $n$ is suppressed by higher loop effects and/or a strong form factor cutoff~\cite{mv94,vg,son}. Recently there has been a certain revival of interest both to the methods developed in the course of those studies, in particular in connection with the possibility of double Higgs boson production at LHC~\cite{lv},  and to the idea of an observably large cross section for production of multiple weak interaction bosons at multi TeV energies~\cite{khoze,jk,ks}.  The latter idea is being discussed using the past results found in simplified models. In particular, for a purely multi Higgs boson process $1 \to n$, where one virtual Higgs particle  produces $n$ bosons, the behavior of the rate ${\cal R}$ in a theory of the scalar field was shown~\cite{lrst,lst} to obey the scaling behavior ${\cal R} \sim \exp[n \, F(n \lambda, \epsilon)]$ in the limit $n \to \infty$, $\lambda \to 0$, $n \lambda$-fixed, and $\epsilon$ being the kinetic energy per final particle. The amplitude for this process at $\epsilon = 0$, i.e. exactly at the threshold for $n$ scalar bosons, is in fact known explicitly at the tree level~\cite{mv92,akp,brown} as well as with the one loop correction generated by the scalar field self interaction~\cite{mv92l,smith}:
\be
A_n= n! \, (2 v)^{1-n} \left [ 1+ n (n-1) \, { \sqrt{3} \, \lambda \over 8 \pi} + O(\lambda^2) \right]~,
\label{an}
\ee
where $v$ is the (classical) vacuum mean value of the scalar field, related to the coupling $\lambda$ and the scalar mass $\mu$ as $\mu^2= 2 \lambda \, v^2$. Clearly, this expression is in agreement with the scaling behavior, once the loop correction is exponentiated~\cite{lrst}. 

It should be pointed out however that the scaling behavior is only applicable in a theory of one bosonic field with one coupling. In a theory where the considered scalar field interacts with heavy particles the scaling behavior is in fact not sustainable and is generally broken by loops with heavy particles.
Indeed, if the scalar field four-momentum is neglected, integrating out heavy particles produces an effective Lagrangian with powers of the considered bosonic field $\phi$: $\xi_k \, \phi^k$, and where $\xi_k$ are the corresponding couplings. One can readily verify that inserting such vertex in interaction between $n$ final particles results in a correction with relative value $n^{k-2} \xi_k$. Clearly, the approximation where the four-momentum of the scalar particles can be neglected is not applicable if the total mass of a cluster of $~k$ scalar bosons is larger than the mass $M$ of the particle in the loop. Thus the power of $n$ in the relative correction due to the loop is of order $M/\mu$ and at sufficiently large ratio of the masses becomes larger than two in violation of the scaling law.  In connection with this behavior in the only case of potentially practical interest, i.e. for the actual Higgs field, the effect of the top quark loop certainly merits a detailed consideration. In what follows the correction to the amplitude $A_n$ in Eq.(\ref{an}) generated by a loop with a fermion acquiring all of its mass $m$ from the interaction with the Higgs field is calculated in the limit of large $n$. As expected from the reasoning outlined above the power of $n$ in this correction is determined by the ratio of the masses $r= m/\mu$:
\be
A_n \to A_n \times \left [ 1 + (-1)^{2 r} \, C (r) \, n^{4 r-4} \, \lambda \right] 
\label{anres}
\ee
With the coefficient $ C(r)$ given by Eq.(\ref{corres}) below. The imaginary part of the correction contained in the factor $(-1)^{2 r}$ corresponds to the unitary cut across the fermion loop. This imaginary part vanishes when $2 r$ is integer. This is a consequence of the property of `nullification'~\cite{mv92z} at integer ratio $2 m /\mu$, i.e. of the exact cancellation to zero of all the on-shell amplitudes for fermion-antifermion annihilation into any number of higgs bosons all being at rest.     

One can readily estimate that with $m$ and $\mu$  being the actual top quark and  Higgs boson  masses,  $m/\mu \approx 1.4$, the power of $n$ in the correction is 1.6 and is smaller than two. Thus the purely bosonic correction in Eq.(\ref{an}) formally exceeds the effect of the top loop at sufficiently large $n$. However the coefficient $C(r)$ is actually numerically large,  $(-1)^{2.8} \, C(1.4) \approx - (8.0 - i \, 5.8) \, \sqrt{3}/(8 \pi)$. Thus the bosonic term equals the real part of the contribution of the top loop at $n^{0.4} \approx 8$ i.e.  at $n \approx 180$. Clearly, at such $n$ each of the corrections becomes very large and far beyond any reasonable justification for considering them in the first order. It thus appears impossible, at the present level of understanding of multi boson processes, to come to any conclusions about their phenomenological significance.

Furthermore, it not yet excluded that there exist heavy fermions and bosons that acquire from the Higgs field a larger mass than that of the top quark. Their loops would then generate corrections to the multi Higgs processes with the power of $n$ larger than two, and those contributions would thus explicitly violate the scaling behavior  and be potentially very important.

The rest of this paper contains a somewhat detailed outline of the calculation of the fermion loop contribution to the amplitude $A_n$, and a result for the effect of the loop with a massive scalar. These calculations employ the approach~\cite{brown} in which the background field in the Euclidean time $\phi(\tau)$ describes the generating function for the amplitudes $A_n \equiv <n|\phi(0)|0>$. A detailed derivation and the description of this approach for the tree level and one loop amplitudes can be found elsewhere~\cite{brown,mv92,mv92l,smith}. Here I briefly describe the steps in the calculation.  Let $\phi_0(\tau)= v + \sigma_0(\tau)$ be the solution to the classical field equation for the boson field approaching the vacuum $v$ at Euclidean infinity, $\tau \to +\infty$. The deviation $\sigma_0$ approaches zero as a series in the exponent $u=\exp(-\mu \tau)$: $\sigma_0 = \sum_{n=1}^\infty c_n u^n$. The tree level amplitudes are then expressed through the coefficients of the expansion as $A_n = n! c_n/(c_1)^n$. The division by the power of the coefficient $c_1$ (the norm of the one-particle state) ensures that the so derived amplitudes do not depend on a shift of the solution $\phi(\tau)$ by a finite time. The quantum loops generate corrections to the background field, so that $\sigma= \sigma_0+ \sigma_1 + \ldots$, and the full quantum amplitude is calculated from the full $\sigma$ as
\be
A_n=  \left . {d^n \over du^n} \sigma \,  \left ( {d \over du} \sigma \right )^{-n} \right |_{u=0}~.
\label{genf}
\ee 

The Largangian for the Higgs scalar plus the top quarks can be written in terms of the real Higgs field component $\phi(x)=v + \sigma(x) $ as
\be
L={1 \over 2} (\partial_\mu \phi)^2 + i \, (\bar t \gamma^\mu \partial_\mu t) - {\lambda \over 4} \, \left ( \phi^2 - v^2 \right )^2 - {m \over v} \, \phi \, (\bar t t)~,
\label{lagr}
\ee
which describes the Higgs scalars with mass $\mu = \sqrt{2 \lambda } \, v$ and the top quarks with the mass $m$.
The spatially uniform classical background field in the Euclidean time $\tau$ has the form 
\be
\phi_0 = v + \sigma_0 = v \, \tanh {\mu \tau \over 2} = v \, { 1-u\over 1+u}~. 
\label{phi0}
\ee
Clearly, this expression generates the tree level amplitudes $A_n$ described by the corresponding part of Eq.(\ref{an}). The classical field has a singularity (in the complex plane of $\tau$) at $u=-1$, and the quantum corrections generally develop a singularity at the same point. The asymptotic at high order behavior of the coefficients in the Taylor series in $u$ for $\sigma$ is determined by the behavior at this singularity. Thus, according to Eq.(\ref{genf}) the calculation of the asymptotic in $n$ behavior of the corrections amounts to determining the corresponding correction to $\sigma(\tau)$ near the singularity. 

\begin{figure}[ht]
\begin{center}
 \leavevmode
    \epsfxsize=8cm
    \epsfbox{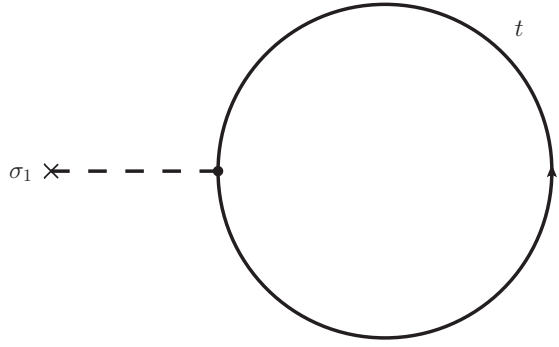}
    \caption{The tadpole graph for the correction $\sigma_1(\tau)$ to the background scalar field. The propagators for the fermion (solid) and the scalar (dashed) are the exact Green's functions in the classical background field $\sigma_0(\tau)$.}  
\end{center}
\end{figure}

The correction $\sigma_1(\tau)$ to the profile of the scalar field due to the top quark loop is described by the `tadpole' graph of Fig.~1 and is a solution to the equation
\be
\left ( {d ^2 \over d \tau^2} - \mu^2 - 3 \mu^2 {\sigma_0 \over v} - {3 \mu^2 \over 2} \, { \sigma_0^2 \over v^2} \right ) \, \sigma_1 = {m \over v} \, < \bar t t >~.
\label{s1eq}
\ee
The source term in this equation is generated by the loop and can be written in terms of the Euclidean space fermion Green's function $G(\tau, \vec x; \tau' , \vec x')$ at coinciding points as 
\be
\eta(\tau) \equiv {m \over v} \, < \bar t t > = - {m \over v} \, N_c \, {\rm Tr} \, G(\tau, 0 ; \tau, 0)~,
\label{etag}
\ee
where $N_c = 3$ is the number of colors, and the equation for the Green's function reads as
\be
\left [ \gamma_0 \partial_\tau +i \, (\vec \gamma \cdot \vec \partial_{x}) - {m \over v} \, \phi_0 (\tau)  \right ] G(\tau, \vec x; \tau' , \vec x') = -\delta(\tau - \tau') \, \delta^{3} (\vec x - \vec x')~.
\label{eqfg}
\ee 
Due to the spatial uniformity of the background field one can make use of the Green's function $G_p(\tau, \tau')$ in the mixed representation:
\be
G(\tau, \vec x; \tau' , \vec x') = \int G_p(\tau , \tau') \, e^{i \vec p \cdot (\vec x - \vec x')}  \, {d^3 p \over (2 \pi)^3}~.
\label{gp}
\ee
The latter  function can be sought for in the form 
\be
G_p(\tau, \tau') = \left [ \gamma_0 \partial_\tau - (\vec \gamma \cdot \vec p) + {m \over v} \, \phi_0 (\tau)  \right ] \, D_p(\tau, \tau')~,
\label{gpfp}
\ee
with the equation, following from Eq.(\ref{eqfg}) for the matrix function $D_p$:
\be
\left [- {d^2 \over d \tau^2} + \vec p^{\, 2} + {m^2 \over v^2} \, \phi_0^2 - {m \over v} \, \gamma_0 \, \left ({d \phi_0 \over d \tau} \right ) \right ] D_p(\tau, \tau') = \delta  (\tau - \tau')~.
\label{eqf}
\ee
Using the standard representation for $\gamma_0$: $\gamma_0 = {\rm diag}(1,-1)$ in the $2 \times 2$ matrix notation, and writing in the same notation the matrix $D_p$ in the form $D_p= {\rm diag}(A_p,B_p)$, one finds that equations for the Green's function reduce to those for the scalar functions $A_p(\tau, \tau')$ and $B_p(\tau, \tau')$ in the form
\be
(P P^\dagger +\vec p^{\, 2})  A_p = \delta(\tau - \tau')~,~~~~~~(P^\dagger P +\vec p^{\,2})  B_p = \delta(\tau - \tau')~,
\label{abeq}
\ee  
where the operators $P$ and $P^\dagger$ are defined as 
\be
P=- {d \over d \tau} + {m \over v} \, \phi_0 (\tau)~,~~~~~~~P^\dagger= {d \over d \tau} + {m \over v} \, \phi_0 (\tau)~.
\label{op}
\ee
The source term in Eq.(\ref{etag}) is then expressed through $A_p$ and $B_p$ by the formula
\be
\eta(\tau) = - 2 \, {m \over v} \int {d^3 p \over (2 \pi)^3} \left . \left [ P^\dagger A_p(\tau, \tau') + P B_p(\tau, \tau') \right ] \right |_{\tau' = \tau}~.
\label{etaab}
\ee
 
The differential operators in the equations (\ref{abeq}) are of the familiar P\"oschl-Teller type:
\be
{4 \over \mu^2} \left ( P^\dagger P+ \vec p^{\,2} \right ) = -{d^2 \over d y^2} + w^2 - {s (s+1) \over \cosh^2 y}~,~~~~~~{4 \over \mu^2} \left ( P P^\dagger + \vec p^{\,2} \right )  = -{d^2 \over d y^2} + w^2 - {(s+1)(s+2) \over \cosh^2 y}~,
\label{pto}
\ee
where the following notation is used: $y = \mu \tau/2$, $w= 2 \omega / \mu$ with $\omega^2 = m^2 + \vec p^{\,2}$, and
\be
s = {2 m \over \mu} -1 ~.
\label{defs}
\ee
Thus the zero energy solutions to the equations can be readily written. The regular at  $\tau \to +\infty$ solutions to the equations $(P P^\dagger + \vec p^{\,2}) \, a_+(\tau) = 0$ and $(P^\dagger P+ \vec p^{\,2}) \, b_+(\tau) =0$ are expressed in terms of the standard hypergeometric function $_2F_1$ as follows
\bea
&&a_+(\tau)= u^{w/2} \, _2F_1 \left ( s+2, -s - 1; w+1; {u \over 1+u} \right )~, \nonumber \\ 
&&b_+(\tau)= u^{w/2} \, _2F_1 \left ( s+1, -s; w+1; {u \over 1+u} \right )~.
\label{abp}
\eea
These solutions are obviously related by the formulas
\be
P^\dagger \, a_+ = - (\omega - m) \, b_+~, ~~~~~~P \, b_+ = (\omega + m) \, a_+~.
\label{abr}
\ee 
The solutions $a_-(\tau)$ and $b_-(\tau)$ regular at $\tau \to - \infty$ are obtained by making in the functions in Eq.(\ref{abp}) the replacement $u \to 1/u$:
\bea
&&a_-(\tau)= u^{-w/2} \, _2F_1 \left ( s+2, -s - 1; w+1; {1 \over 1+u} \right )~, \nonumber \\ 
&&b_-(\tau)= u^{-w/2} \, _2F_1 \left ( s+1, -s; w+1; {1 \over 1+u} \right )~.
\label{abm}
\eea

The Green's functions in Eq.(\ref{abeq}) are then found as 
\bea
&&A_p(\tau, \tau') = {1 \over W_a} \, \left [  a_+(\tau) a_-(\tau') \, \theta (\tau - \tau') +   a_+(\tau') a_-(\tau) \, \theta (\tau' - \tau) \right ]~, \nonumber \\
&&B_p(\tau, \tau') = {1 \over W_b} \, \left [  b_+(\tau) b_-(\tau') \, \theta (\tau - \tau') +   b_+(\tau') b_-(\tau) \, \theta (\tau' - \tau) \right ]~,
\label{gfab}
\eea
where $W_a$ and $W_b$ are the corresponding Wronskians:
\bea
&&W_a=a_+ {d a_- \over d \tau}- a_- {d a_+ \over d \tau} = 2 \omega \, { \Gamma(w+1) \Gamma(w) \over \Gamma(w-s-1) \Gamma(w+s+2)}~, \nonumber \\
&&W_b=b_+ {d b_- \over d \tau}- b_- {d b_+ \over d \tau} = 2 \omega \, { \Gamma(w+1) \Gamma(w) \over \Gamma(w-s) \Gamma(w+s+1)}~,
\label{wrsk}
\eea
as can be readily found by using the well known relation~\cite{abra} between the hypergeometric functions at $z$ and $1-z$ to find the leading (growing) asymptotic behavior of the functions $a_-$ and $b_-$ at $\tau \to + \infty$.

Using the expressions (\ref{gfab}) and (\ref{wrsk}), and also the relations (\ref{abr}) the integrand in Eq.(\ref{etaab}) can be 
found in the form
\bea
&&\! \! \! \! \! \! \! \! \! \! \! \! \! \! \! \left . \left [ P^\dagger A_p(\tau, \tau') + P B_p(\tau, \tau') \right ] \right |_{\tau' = \tau} = {\Gamma(w-s) \Gamma(w+s+1) \over 2 \omega \, \Gamma(w+1) \Gamma(w)} \left [ (\omega+m) \, a_+(\tau) b_-(\tau) - \right . \nonumber \\ 
&&\! \! \! \! \! \! \! \! \! \! \! \! \! \! \! \! \! \! \! \! \! \left . {(\omega-m) (w+s+1) \over (w-s-1) } \, b_+(\tau) a_-(\tau) \right] = {\Gamma(w-s) \Gamma(w+s+2) \over 2  \Gamma^2(w+1)} \, \left [ a_+(\tau) b_ -(\tau) - b_+(\tau) a_-(\tau) \right]~.
\label{inex}
\eea

The leading behavior of the latter expression at the singularity of the background field at $\tau \to i \pi/\mu$, or equivalently at $u \to -1$ can be found by using the standard formula~\cite{abra} for relation between the hypergeometric functions at $z$ and $1/z$. This leading term is given by
\be
\left . (a_+ b_- - a_-b_+) \right |_{u \to -1} = 4 (-1)^s  {\Gamma^2(w+1) \Gamma(2s+1) \Gamma(2s+2) \over \Gamma^2(s+1) \Gamma(w+s+1) \Gamma(w+s+2)} \, (1+u)^{-2s-1} \, \left [ 1 + O(1+u) \right ]~.
\label{sing} 
\ee
Combining this expression with that in Eq.(\ref{inex}) one finds the leading singularity in the source term in Eq.(\ref{etag}) in the form
\be
\left . \eta \right |_{u \to -1} =  - (-1)^s \, N_c \, (s+1) \, {\Gamma(2s+1) \Gamma(2s +2) \over \Gamma^2(s+1)} \, I(s) \, {\mu^4 \over 8 \pi^2 \, v}  \, (1+u)^{-2s-1} \, \left [ 1 + O(1+u) \right ]~,
\label{etares}
\ee
with $I(s)$ defined as
\be
I(s) =\int_{w=s+1}^\infty \, {\Gamma(w-s) \over \Gamma(w+s+1)} \, w \, \sqrt{w^2 - (s+1)^2} \, dw~.
\label{iofs}
\ee

The leading singularity at $u \to -1$ of the correction $\sigma_1(\tau)$ to the background scalar field is thus found by retaining the most singular term proportional to $\sigma_0^2$ in the l.h.s. of the equation (\ref{s1eq}) and using the expression (\ref{etares}) for the source term. In this way one readily finds
\be
\left . \sigma_1 \right |_{u \to -1} =  - (-1)^s \, N_c \,  {\Gamma(2s+1) \Gamma(2s +2) \over (2s-3) \Gamma^2(s+1)} \, I(s) \, {\mu^2 \over 16 \pi^2 \, v}  \, (1+u)^{-2s+1} \, \left [ 1 + O(1+u) \right ]~ ,
\label{s1res}
\ee 
and thus determines the leading at large $n$ behavior of the ratio of the $n$-th derivatives of $\sigma_1$ and $\sigma_0$ [Eq.(\ref{phi0})] with respect to $u$ at $u=0$:
\be
\left . {(d/du)^n \sigma_1 \over (d/du)^n \sigma_0} \right |_{u=0} = - (-1)^s N_c \, {(2 s -1) \, s \, \Gamma(2s+2) \over (2s -3 ) \, \Gamma^2 (s+1) } \, I(s) \, n^{2s-2} \, {\lambda \over 8 \pi^2}~.
\label{corres}
\ee
Given that the expression for the field $\sigma = \sigma_0+\sigma_1$ is the generating function for the amplitudes $A_n$ with the one-loop correction, one arrives at the formula (\ref{anres}) with the coefficient $C(r)$ explicitly given by
\be
C(r) = {N_c \over 8 \pi^2} \,  {(2 r -1) \, (4r-3) \, \Gamma(4 r) \over (4 r -5) \, \Gamma^2 (2 r) } \, I (2r -1)~.
\label{cofr}
\ee

One can also consider a model where the Higgs field interacts with a heavy scalar $X$. The loop correction to a multi Higgs production is described by the quadratic in $X$ part of the Lagrangian. Assuming $X$ to be real,  this part can be written as
\be
L_X= {1 \over 2} \, \left ( \partial_\mu X \right )^2 - {M^2 \over 2 } \, X^2 - {\lambda_X \over 2} \, \phi^2 \, X^2~,
\label{lx2}
\ee
where $\lambda_X$ is the dimensionless coupling between the Higgs field and the $X$, and the mass $M_X$ of the $X$ particle in the Higgs vacuum is given by $M_X^2 = M^2 + \lambda_X \, v^2$.   A simple calculation along the same lines as described above for a fermion, yields the asymptotic behavior of the loop correction to the amplitudes $A_n$ which behavior sets in at $n$ larger than $M_X/\mu$:
\be
A_n \to A_n \left [ 1 + (-1)^{s_X} \, C_X \left ( {M \over \mu}, s_X \right ) \, n^{2 s_X -2} \, \lambda \right ]~,
\label{anx}
\ee
where the (positive) index $s_X$ is related to the ratio of the scalar couplings as
\be
s_X (s_X+1) = 2 \, {\lambda_X \over \lambda}~,
\label{sx}
\ee
and the coefficient function is
\be
C_X= {1 \over 32 \pi^2} \, { s_X \, (2 s_X-1) \, \Gamma(2s_X+1) \over (2s_X-3) \, \Gamma(s_X) \, \Gamma(s_X+1)} \int_{2 M_X \over \mu}^\infty \, {\Gamma(w-s_X) \over \Gamma(w+s_X+1)} \, w \, \sqrt{w^2 - 4 M_X^2/\mu^2} \, dw~. 
\label{cx}
\ee
In the limit of a very heavy $X$, $M_X \gg \mu$, the integral can be approximated analytically, and the expression for $C_X$ takes the form
\be
C_X \approx {1 \over 32 \pi^2} \, {2 s_X-1  \over  (s_X-1) \, (2 s_X-3)} \, \left ({\mu \over M_X} \right )^{2 s_X-2}~.
\label{cxlim}
\ee
The correction from the scalar loop is real if the index $s_X$ is integer, in agreement with the nullification of all the on-shell threshold amplitudes for the production of multi Higgs states by two scalar $X$ bosons~\cite{mv92z}.

In summary. The amplitude $<n|\,\phi(0)|\, 0>$ for the  production of $n$ static Higgs bosons by one virtual field receives loop corrections that are rapidly growing with $n$. The calculated here corrections from a loop of a heavy fermion, e.g. the top quark, or a heavy scalar generally break the scaling behavior of the corrections with  $n$ and $\lambda$ that was inferred for models of one field. Numerically the corrections due to the top quark loop are large and in practice make it impossible to arrive at any conclusions regarding phenomenological significance of the discussed multi Higgs processes.

This work is supported in part by U.S. Department of Energy Grant No.\ DE-SC0011842.

\end{document}